\documentclass[journal=jacsat,manuscript=article]{achemso}

\usepackage{chemformula} % Formula subscripts using \ch{}
\usepackage[T1]{fontenc} % Use modern font encodings
\usepackage{microtype}
\usepackage{graphicx}
\usepackage{array}
\usepackage{colortbl}
\usepackage{epstopdf}
\usepackage{booktabs}
\usepackage{url}
\usepackage{adjustbox}
\arrayrulecolor{black}
\usepackage{multirow}
\usepackage{tabularray}
\usepackage{soul}
\usepackage{makecell}

\author{Jiahui Zhang}
\affiliation{MOE Key Laboratory of Environmental Theoretical Chemistry, Guangdong Provincial Key Laboratory of Chemical Pollution and Environmental Safety, SCNU Environmental Research Institute, School of Environment, South China Normal University, Guangzhou 510006, P. R. China.}
\author{Yifei Zhu}
\affiliation{MOE Key Laboratory of Environmental Theoretical Chemistry, Guangdong Provincial Key Laboratory of Chemical Pollution and Environmental Safety, SCNU Environmental Research Institute, School of Environment, South China Normal University, Guangzhou 510006, P. R. China.}
\author{Chuqiao Feng}
\affiliation{MOE Key Laboratory of Environmental Theoretical Chemistry, Guangdong Provincial Key Laboratory of Chemical Pollution and Environmental Safety, SCNU Environmental Research Institute, School of Environment, South China Normal University, Guangzhou 510006, P. R. China.}
\author{Yingjin Ma}
\affiliation{Computer Network Information Center, Chinese Academy of Sciences, Beijing 100190, China}
\author{Chao Xu}
\affiliation{MOE Key Laboratory of Environmental Theoretical Chemistry, Guangdong Provincial Key Laboratory of Chemical Pollution and Environmental Safety, SCNU Environmental Research Institute, School of Environment, South China Normal University, Guangzhou 510006, P. R. China.}
\email{chaoxu@m.scnu.edu.cn}
\author{Zhenggang Lan}
\affiliation{MOE Key Laboratory of Environmental Theoretical Chemistry, Guangdong Provincial Key Laboratory of Chemical Pollution and Environmental Safety, SCNU Environmental Research Institute, School of Environment, South China Normal University, Guangzhou 510006, P. R. China.}
\email{zhenggang.lan@m.scnu.edu.cn}

\title{A quantum chemistry dataset containing $S_0$-$S_1$ conical-intersection structures of 259k molecules}

\begin{document}
\graphicspath{{figure/}}

\begin{abstract}

Conical intersections are key to photoinduced reactions, but comprehensive datasets of their structures remain rare. To address this gap, we built the QCDGE-CI dataset, which contains ground-state and minimum-energy conical-intersection structures for more than 259k small molecules with up to ten heavy atoms (C, N, O, or F). The minimum-energy conical-intersection geometries were computed using the semi-empirical OM2/MRCI method and evaluated through systematic benchmarks with an enlarged active space and CASSCF calculations. These benchmarks support the reliability of the geometries. Cheminformatics analyses further reveal the broad chemical diversity covered by the dataset. The dataset is designed to support machine-learning and data-driven studies of excited-state chemistry, including nonadiabatic processes.

\end{abstract}

\section{Introduction}

Photophysics and photochemistry are central to many processes in daily life, as sunlight drives numerous critical events on Earth.~\cite{Domcke2004, Domcke2011, Domcke2012, Crespo-Otero2018, Barbatti2015, Curchod2018, Matsika2011, Segatta2019} When a molecule absorbs light, it undergoes an electronic transition to an excited state. On the excited-state potential energy surface, nuclear motion can bring the system to regions where two electronic states of the same spin become equal in energy. In these regions, the two coupled surfaces form a characteristic double cone, which is why they are called conical intersections (CIs).~\cite{Domcke2004, Domcke2011, Domcke2012, Matsika2011, Crespo-Otero2018, Curchod2018} Near a CI, the molecule can switch between electronic states without emitting light. This process, known as nonadiabatic dynamics, plays a crucial role in photoinduced reactions and often determines the outcome of photochemical processes. CIs are therefore frequently described as the "funnels" of photochemistry.~\cite{Domcke2004, Domcke2011, Bonacic1987} Understanding these intersections is essential for explaining how photochemical reactions proceed.

The actual dynamics after photoexcitation can be complex and depend strongly on the system. 
In some cases, the system relaxes to an excited-state minimum before reaching a CI. 
In others, the CI is accessed directly from the Franck-Condon region without passing through a distinct minimum. 
In still other cases, a minimum exists but the CI is too high in energy to be reached, so the molecule cannot decay through that funnel. 
The excited-state minimum itself may also be absent, which leads to other relaxation pathways. 
Given this variety, a clear starting point is needed for systematic CI studies. 
The minimum-energy conical intersection (MECI) on the CI seam provides such a reference~\cite{Levine_arpc_2007, Domcke2004, Domcke2011, Domcke2012, Crespo-Otero2018, Barbatti2015, Curchod2018, Matsika2011, Segatta2019, List_cs_2022, Pieri_cs_2021}. 
As the lowest-energy point on the seam, the MECI gives a representative structure for understanding the CI seam and its associated nonadiabatic dynamics. 
A systematic database of MECIs across many molecular systems would support the exploration of photochemical pathways and the development of data-driven models for excited-state chemistry.

Despite their fundamental importance, CI studies remain highly system-specific. 
Typically, researchers identify CIs with suitable electronic structure methods.~\cite{Domcke2004, Domcke2011} 
They investigate nonadiabatic dynamics with various simulation approaches.~\cite{Domcke2004, Domcke2011, Domcke2012, Matsika2011, Crespo-Otero2018, Curchod2018, mai2018nonadiabatic, SHARC4.0} 
Others also explore the CI seam using specialized algorithms.~\cite{Mori_jctc_2012, Liu_jcp_2025, Lindner_jctc_2019} 
These studies are performed for one molecule at a time, often requiring substantial manual effort and expert knowledge. 
While such detailed work provides deep mechanistic insights, it is hard to scale up. 
As a result, our understanding of CIs across broad chemical spaces remains fragmented and incomplete.

Artificial intelligence can offer a way to address these challenges. 
In recent decades, machine learning (ML) has become increasingly important in chemical research.~\cite{Aldossary2024, Shao2023, Glielmo2021, Dral2021, Westermayr2021, Li2023, li2021computational, bu2023designing, chen2026unified, Hu2018, Lin2022} 
However, ML applications in photochemistry and photophysics are still limited to a few topics, 
such as predicting excitation energies~\cite{Pronobis2018, Srsen2024, Xiao2026, Coxson2026} 
and accelerating excited-state dynamics.~\cite{Hu2018, Chen2018, Dral2018, Li2021, Lin2022, Tang2022} 
ML studies that focus on CI properties across many compounds remain rare. 
The key reason is the lack of a suitable quantum chemistry CI dataset. 
Several ground-state datasets exist~\cite{Pereira2017, Yuan2025, ullahMolecularQuantumChemical2024, Ramakrishnan2014, rupp2012qm7, Montavon2013qm7b,
axelrodGEOMEnergyannotatedMolecular2022, smithANI1ExtensibleNeural2017, bowmanMD17DatasetsPerspective2022, pinheirojrWS22DatabaseWigner2023, zhangVIB5DatabaseAccurate2022}, 
and some provide preliminary excited-state information.~\cite{Zhu2024, ramakrishnanElectronicSpectraTDDFT2015, zouDeepLearningModel2023, liangQMsymexUpdateQMsym2020, nakataPubChemQCProjectLargeScale2017, pengmeiMD17ReactiveXxMD2024, lupopasiniTwoExcitedstateDatasets2023} 
However, comprehensive datasets that include MECI geometries and energies across broad chemical spaces are still missing. 
As a consequence, ML models for CI prediction cannot be trained and validated at scale across diverse molecular systems.

To fill this gap, we built the QCDGE-CI dataset, which provides MECI structures and ground-state geometries for over 259k small molecules. 
This is the first large-scale dataset of its kind, with molecular structures derived from the QCDGE dataset.~\cite{Zhu2024} 
We optimized the ground-state geometries at the OM2 level. 
Single-point calculations were then performed at the OM2/MRCI (Orthogonalization Model 2 - Multireference Configuration Interaction) level.~\cite{Weber2000, Koslowski2003, Keal2007} 
MECI geometries and energies were obtained at the same level.~\cite{Keal2007} 
We chose the OM2/MRCI method because it offers a good balance between efficiency and accuracy for CI characterization.~\cite{Keal2007, Nikiforov2014, Domcke2011} 
This choice makes it feasible to process hundreds of thousands of molecules. 
The dataset covers small molecules with up to ten heavy atoms (C, N, O, F). 
It is designed for ML training, statistical analysis, and preliminary screening of photochemical properties.

The main dataset was generated with a fixed (4,4) active space. 
To assess its reliability, we performed two types of benchmark calculations. 
First, we re-optimized all converged CI structures using an enlarged (8,8) active space at the same OM2/MRCI level. This allowed us to check the consistency of the (4,4) geometries and to detect any systematic biases. 
Second, we re-optimized a representative subset at the Completed-Active-Space Self-Consistent Field  (CASSCF) level~\cite{Malmqvist1989, Olsen2011} using two different active spaces. 
This provides a high-level reference for the semi-empirical method. 
These benchmarks, together with convergence statistics and RMSD analysis~\cite{Kagsch_1976, Kabsch_1978}, give a quantitative basis for assigning confidence levels to different subsets of the dataset. 
We also performed extensive chem-informatics analyses to characterize the chemical space coverage. 
These include elemental composition, compound types, ring distributions, functional groups, molecular shapes, bond lengths, and energy distributions. 
This multi-faceted validation ensures that the QCDGE-CI dataset is not only large in scale,
but also well-characterized in terms of accuracy, diversity, and applicability.

Among all types of CIs, the ones between the ground state ($S_0$) and the first excited state ($S_1$) are especially important. They are involved in many nonadiabatic processes, including photoisomerization and the photostability of biological molecules.~\cite{Domcke2004, Domcke2011, Domcke2012, Bonacic1987, Crespo-Otero2018, Barbatti2015, Curchod2018, Matsika2011} 
These $S_0$-$S_1$ crossings are also computationally more tractable than those involving higher excited states. The OM2/MRCI method also gives better accuracy for this lowest-lying crossing.~\cite{Keal2007, Nikiforov2014, Domcke2011} For these reasons, the present dataset focuses exclusively on $S_0$-$S_1$ CIs.

The QCDGE-CI dataset is intended for qualitative to semi-quantitative applications. 
These include ML model training, large-scale pattern analysis, and preliminary screening of CI properties. 
It is not designed to replace high-level multi-reference calculations for individual systems.~\cite{Szabo1989, Domcke2004} 
The OM2/MRCI method has inherent approximations, and users should be aware of them. 
Nevertheless, we believe this dataset will open new opportunities for large-scale CI exploration and will be useful for a wide range of photochemical and photophysical studies.

\section{Computational Methods and Implementation Details.}

\subsection{Selection of Electronic Structure Methods}

A key goal of this work is the large-scale optimization of $S_0$-$S_1$ conical intersections across many small molecules containing H, C, N, O, and F atoms. 
This task has four main requirements for the electronic structure method.
The electronic structure method must correctly describe the degeneracy and topology of conical intersections. At the same time, it needs to be efficient enough for hundreds of thousands of molecules and robust across diverse systems. It must also provide energy gradients and nonadiabatic coupling vectors for geometry optimizations.

Single-determinant methods like Hartree-Fock (HF) and density functional theory (DFT) are widely used for ground-state structures and reactions because they are computationally efficient. 
However, they cannot describe excited states. 
To address this, methods such as configuration interaction singles (CIS) and time-dependent DFT (TDDFT)~\cite{Marques2012} were developed for excited-state calculations, including absorption and emission spectra. Nevertheless, these methods still fail for $S_0$-$S_1$ 
conical intersections. 
In particular, TDDFT cannot correctly describe the potential-energy-surface topology near a crossing that involves the ground state.~\cite{levine2006conical} 
Consequently, we did not consider single-reference methods further.

Multi-reference methods like CASSCF,~\cite{Malmqvist1989, Olsen2011} and CASPT2~\cite{finley1998multi} treat molecular excited states rigorously. 
They handle the multi-configurational wavefunction at degeneracies correctly and can give accurate CI geometries, energies, gradients, and couplings. 
In principle, these approaches describe CI topologies well. 
In practice, however, they are impractical for a dataset of this scale. 
The main obstacles are high computational cost, frequent convergence failures, and strong dependence on active-space selection. 
Taking CASSCF as an example, its cost grows rapidly with basis-set size and active-space dimension. 
Convergence and final results also depend strongly on the active-space choice, 
which is hard to automate for thousands of diverse molecules. CASPT2 are even more costly. 
These high-level methods are therefore not feasible for large-scale construction, though they remain useful for benchmarking on representative subsets.

Another alternative is spin-flip (SF) methods~\cite{Shao_jcp_2003}, especially within the TDDFT framework~\cite{TDDFT_2026}. 
These methods use a triplet reference and generate singlet states through spin-flipping excitations. 
This allows them to describe both ground-state and excited-state surfaces, including their crossings, in a single framework. 
However, standard SF-TDDFT methods without spin restriction can suffer from spin contamination.~\cite{lee2018eliminating} 
Spin-restricted variants reduce this problem,~\cite{lee2019efficient} 
but they have not been extensively tested for large-scale use. 
More fundamentally, SF-TDDFT is based on single excitations from a triplet reference. 
The excited-state wavefunctions of some systems contain configurations that cannot be generated this way. 
The method is also computationally demanding for very large-scale applications. 
Although SF-TDDFT has worked well in some cases, 
it has not been widely adopted for large-scale studies across diverse molecules. 
Given our goal of building a broad-coverage dataset, we chose a more established approach, described below.

The OM2/MRCI approach~\cite{Weber2000,koslowski2003implementation, Keal2007} offers a good balance of efficiency and accuracy for excited-state problems, especially for CI optimization. 
It also provides energy gradients and nonadiabatic coupling vectors, which are needed to locate and characterize CIs. OM2~\cite{Weber2000} is a semiempirical method that improves on the standard MNDO model by treating bound and unbound orbitals more evenly. 
Combined with MRCI, it has been widely used to study excited-state dynamics and photochemical mechanisms.~\cite{koslowski2003implementation, Domcke2011} 
Many previous studies have applied OM2/MRCI to CI problems and found that it gives reasonable results for a broad range of molecules.~\cite{Weber2000, koslowski2003implementation, Keal2007, kang2023fly, ma2023simultaneously, dral2016semiempirical} 
OM2/MRCI results are not strongly dependent on the active-space choice, as long as the relevant orbitals are included. This robustness comes in part from the fact that the semiempirical parameters are fitted to experimental data, so dynamic correlation is effectively incorporated at the orbital generation stage.

Based on these considerations, we adopt a hybrid strategy to build the dataset. 
The main dataset is generated with OM2/MRCI using the (4,4) active space, 
which gives the efficiency and scalability needed for over 259k molecules. 

The choice of the (4,4) active space is motivated by the electronic structure near the CI. At an $S_0$-$S_1$ CI, two electronic states become nearly degenerate. In most cases, a bonding-antibonding orbital pair lies close in energy and dominates the electronic structure. In this region, the electronic states are mixtures of the closed-shell ground state and singly excited configurations. The structures before and after the CI may involve a different electronic configuration from that of the closed-shell ground state. This requires at least one further molecular orbital distinct from the bonding-antibonding pair. Together with two electrons and two orbitals involved in the excitation and de-excitation process, this naturally defines a (4,4) active space. This (4,4) choice represents the minimal practical combination that allows the three reference configurations to contribute and provides a balanced description. We therefore used the (4,4) active space for all single-point calculations and CI optimizations.

To access the robustness of the (4,4) results, all converged CI structures are re-optimized with an enlarged (8,8) active space at the same OM2/MRCI level, and the two sets of geometries are subsequently compared. 
In addition, the molecules are classified into eight groups according to their structural features for systematic chemical diversity analysis. On the basis of these groups, a representative subset is selected by randomly sampling structures from each group for CASSCF re-optimization. The accuracy of the OM2/MRCI geometries is then evaluated against the CASSCF results. Two different active spaces are used in the CASSCF calculations to examine the sensitivity of the results. 
The computational details are given in the following subsections.

\subsection{Construction of the CI Dataset}

\begin{figure}[!htbp]
    \centering
    \includegraphics[width=1 \linewidth]{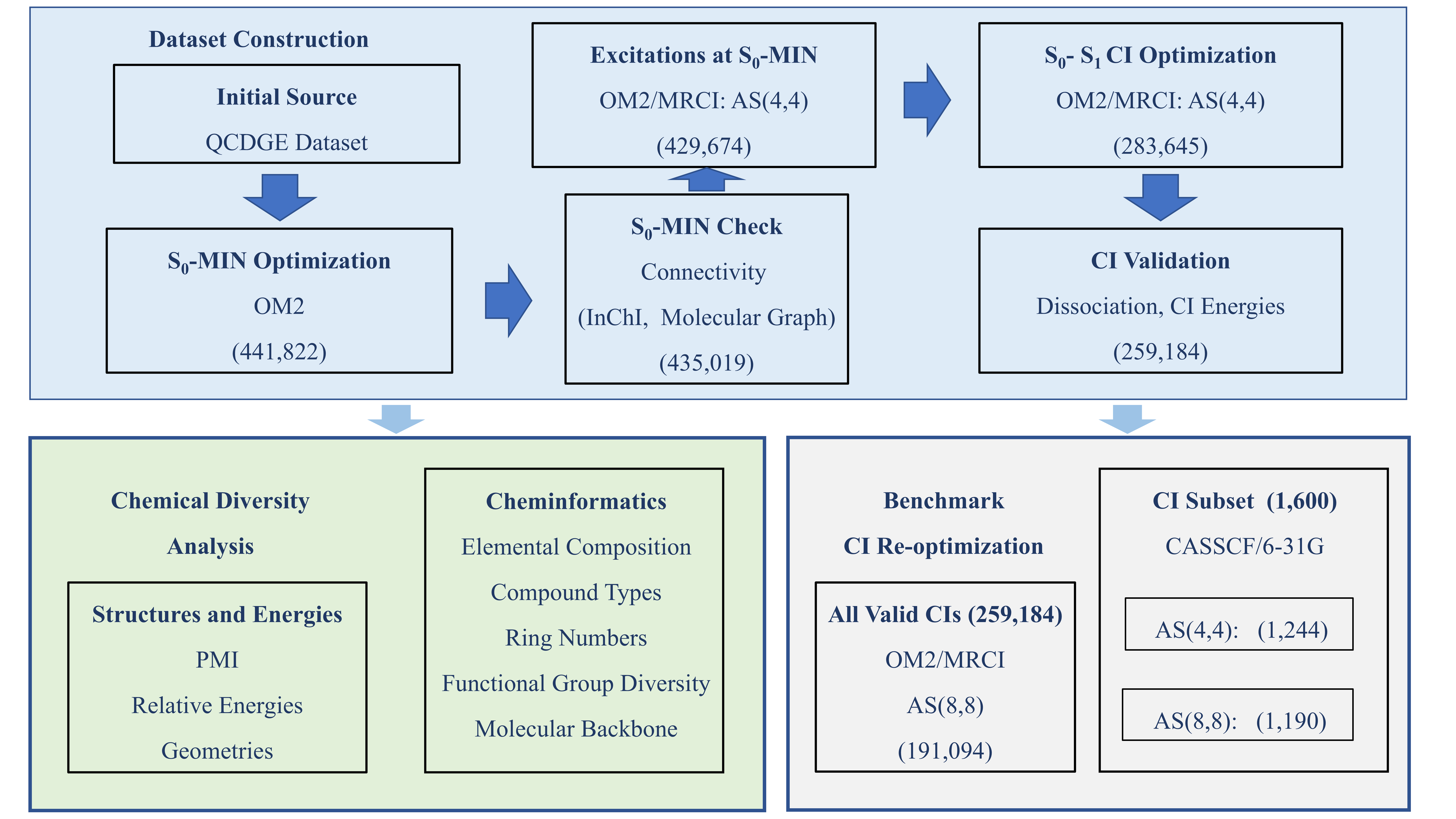}
    \caption{Schematic overview of the dataset construction, chemical diversity analysis, and benchmark selection workflow. The dataset is constructed through geometry optimization, connectivity checks, single-point calculations, CI geometry optimization, and CI validation. The resulting structures are further characterized in terms of chemical diversity, followed by benchmarking of CI geometries using larger active spaces and the CASSCF method.}
    \label{fig:workflow}
\end{figure}

As shown in the dataset-construction stage of the overall workflow in Figure~\ref{fig:workflow}, the QCDGE-CI dataset was constructed through five steps: ground-state optimization at the OM2 level, geometry checking of the optimized ground-state structures, single-point excited-state calculations at the OM2/MRCI level, CI optimization at the same level, and validation of the resulting CI structures.

\subsubsection{Step 1: Ground-state geometry optimizations}

The initial geometries came from the QCDGE dataset~\cite{Zhu2024}, 
in which ground-state optimizations were optimized at the DFT/B3LYP/6-31G(d)/BJD3 level.~\cite{grimme2011effect} 
To maintain consistency with the subsequent CI calculations, we re-optimized all structures at the OM2 level, using the MNDO package.~\cite{MNDO2019}
This procedure provides a consistent electronic-structure description and allows direct comparison of relative energies across the dataset. For small molecules, DFT and semiempirical methods generally produce similar ground-state geometries. Given that QCDGE-CI is intended for large-scale screening and machine-learning applications, the OM2-optimized geometries provide a suitable balance between computational efficiency and structural quality for the subsequent CI calculations.

\subsubsection{Step 2: Optimized ground-state geometry check}

To ensure that the ground-state structures from the two theoretical levels have similar connectivity, we designed a validation procedure that balances cost and accuracy (Figure~S1(a)).

InChI (IUPAC International Chemical Identifier) strings~\cite{heller2015inchi} offer a fast way to check structure equivalence. 
When two InChI strings match, the structures have the same connectivity. 
However, this method has a drawback: small torsional changes can produce different InChI strings for the same molecule, leading to false mismatches.

An alternative approach is to compare molecular graphs directly.~\cite{grambowReactantsProductsTransition2020} Molecules are naturally graphs, with atoms as nodes and bonds as edges. 
Graph comparison gives a precise measure of structural equivalence. 
However, applying it to all molecules is too expensive for large datasets.

We therefore used a hybrid two-step procedure. 
First, we compared InChI strings (generated with RDKit 2022.09.5~\cite{RDKit2022}) for all optimized geometries. Identical strings were taken as matching. 
Second, for cases with different InChI strings, we performed an additional graph comparison to check if the structures were truly different.

These filters ensure that the final geometries are consistent with the original DFT results. To confirm this, we also compared the OM2 and DFT structures. The results are presented later.

\subsubsection{Step 3: Single-point calculations}

Starting from the OM2-optimized ground-state geometries, we performed single-point calculations at the OM2/MRCI level to obtain vertical excitation energies. 
The CI optimizations were then carried to locate $S_0$-$S_1$ MECI structures. 

Our MRCI calculations used three reference configurations: the closed-shell ground state, the single HOMO-LUMO excitation, and the double HOMO-LUMO excitation. The MNDO package~\cite{MNDO2019} does not perform orbital optimization during the MRCI calculation. The orbitals used in the MRCI expansion are therefore determined beforehand, making the choice of the initial orbitals important. Following common practice, we generated these orbitals using the restricted open-shell Hartree-Fock (ROHF) method. This approach provides a reasonable description of the electronic wavefunction near CIs. We chose the (4,4) active space for all single-point and CI calculations, as it is a practical compromise for molecules containing only C, N, O, and F.

\subsubsection{Step 4: Conical intersection (CI) optimizations}

We then optimized the $S_0$-$S_1$ MECI structures 
for all geometries using the Lagrange-Newton algorithm.~\cite{Keal2007, Domcke2004} All settings were the same as those used in the ground-state single-point calculations.

For a dataset of this size, conformational sampling before each CI optimization was not possible. We therefore used the OM2-optimized ground-state geometries as the starting points for all MECI optimizations. This choice may have contributed to convergence failures for some molecules, because the ground-state minimum can be far from the CI region on the potential energy surface. This limitation is inherent to the large-scale nature of the dataset.

To reduce convergence problems, we increased the maximum number of SCF iterations when the default limit was not enough. All other convergence thresholds were kept at their default values in the MNDO package.~\cite{MNDO2019}

\subsubsection{Step 5:  CI validations}

After the CI optimizations, we performed two quality checks on the final dataset (Figure~S1(b)).

First, we checked the molecular graphs of the optimized MECI structures. Fragmentation occurred occasionally during the CI optimizations. Dissociated structures can still show state degeneracy due to radical fragments. Such structures are interesting for dissociation dynamics, but they are not the photochemical funnels that are intermediate structures driving nonadiabatic transitions. We therefore removed all geometries that broke into two or more disconnected pieces.

Second, we excluded CI geometries with energies below the corresponding OM2 ground-state minimum. This can happen when a molecule has multiple ground-state minima and the optimization found a local minimum instead of the global one. It can also happen when the CI optimization caused large structural changes that altered the connectivity. In such cases, the energy ordering may not reflect the true relationship between the MECI and the global ground-state minimum. Since exhaustive conformational searches are not feasible at this scale, we removed these geometries to avoid artifacts. These cases are a small fraction of the total dataset.

Locating conical intersections reliably is challenging in quantum chemistry. The complex electronic structure and unique topology near degeneracies make it difficult. The CI validation procedures described here are practical efforts to ensure the chemical reasonableness of the dataset, while acknowledging the limits of large-scale automated calculations.

\subsection{Chemical Space of CI Dataset}

After building the QCDGE-CI dataset, we performed several validation analyses to assess its chemical space coverage.

To evaluate the chemical diversity, we examined the distributions of elemental composition, compound types, and ring numbers. The eight compound types are heterocycles, fused heterocycles, heteroaromatics, heteroacyclic, carbocycles, carboacyclic, fused carbocycles, and aromatics.

We also used the Ertl algorithm~\cite{Ertl2017, schaub2023development} to study the functional group diversity. The original RDKit version only recognizes a limited set of generic functional groups (C, N, O, F). In our previous work,~\cite{Zhu2024} we extended this to identify up to 109 distinct functional groups; further details are provided in that study. We also performed a Murcko scaffold analysis~\cite{bemis1996properties} to examine the molecular backbones. All cheminformatics analyses used RDKit with our extended setup.

To further characterize molecular shape diversity, we performed a principal moments of inertia (PMI) analysis~\cite{Sauer2003} for both ground-state and 
$S_0$-$S_1$ MECI geometries.
This method classifies molecular shapes by their three-dimensional conformation.
With the principal moments of inertia ordered as $I_1 \le I_2 \le I_3$, molecular shapes can be categorized into three distinct types: rod-like ($I_1 \ll I_2 \approx I_3$), disc-like ($I_1 \approx I_2 \ll I_3$), and spherical ($I_1 \approx I_2 \approx I_3$). 
In a rod-like molecule, the mass distribution is stretched along one axis, giving one very small moment and two large, nearly equal moments. 
In a disc-like molecule, the mass is spread in a plane, yielding two smaller, comparable moments and one much larger moment. 
Spherical molecules have mass evenly distributed, with three comparable moments.

We also examined the distributions of geometries and energies. These include bond lengths (with bond-order resolution), the 
$S_1$ vertical excitation energies (VEEs) 
at the ground-state minima, and 
the $S_0$-$S_1$ CI energies 
relative to the corresponding ground-state minima.

\subsection{Benchmark Calculations}

After building the main dataset, we performed benchmark calculations to assess the reliability of the OM2/MRCI(4,4) geometries. We carried out two sets of benchmarks. First, we enlarged the active space to (8,8) and re-optimized all converged CI structures at the same OM2/MRCI level. This allowed us to check the sensitivity of the geometries to the active-space choice. Second, we selected a representative subset and re-optimized them at the CASSCF level with two different active spaces. This provided a high-level reference for the semi-empirical results.

\subsubsection{Benchmark of the CI geometries with a larger active space}

The main dataset was generated with the (4,4) active space. This choice is a practical compromise for large-scale calculations. However, it may not fully capture the electronic structure of all molecules and could cause convergence problems for some systems. To check the effect of active-space size on the geometries, we re-optimized all CI structures from the (4,4) calculations using an enlarged (8,8) active space at the same OM2/MRCI level.

To document the reliability of each MECI geometry, we added two keywords to the dataset. The first, "Conv\_BigAS", shows whether the enlarged-space optimization converged (1) or not (0). The second, "RMSD\_BigAS", gives the RMSD between the (4,4) and (8,8) geometries for each molecule. For cases that did not converge, "RMSD\_BigAS" is set to -1.

\subsubsection{Benchmark of the CI geometries at the CASSCF level}

The OM2/MRCI method is efficient for large-scale work, but we also need to check its accuracy against a high-level method. Among the multi-reference approaches that can describe $S_0$-$S_1$ CIs, CASSCF is the most practical for benchmarking. Still, re-optimizing all CI geometries at the CASSCF level is too expensive. We therefore used a sampling strategy.

We divided the molecules into eight groups based on their structural features. From each group, we randomly picked 200 CI geometries and re-optimized them at the CASSCF level. To test the effect of active-space choice, we used two different active spaces, (4,4) and (8,8), in these optimizations. In total, we ran 3,200 CASSCF jobs.

To document the CASSCF benchmark results, we introduced several keywords in the dataset. The fields "Conv\_CASSCF\_AS1" and "Conv\_CASSCF\_AS2" indicate whether the CI optimization converged with each of the two active spaces (1 for converged, 0 for not converged). For molecules where at least one of the two CASSCF calculations converged, the RMSD between the OM2/MRCI(4,4) geometry and the corresponding CASSCF geometry was computed and stored in the "RMSD\_CASSCF" field. When both active spaces converged, we used the geometry with the smaller RMSD relative to the OM2/MRCI(4,4) structure as the reference, and the value of the other was not retained. If neither CASSCF calculation converged, "RMSD\_CASSCF" was set to -1. For molecules not selected in the benchmark subset, this field was set to -2.

Our CASSCF benchmarks focus on structural comparisons rather than energetic ones, because CASSCF lacks dynamic correlation and its energies can deviate significantly from higher-level methods such as CASPT2 or MRCI. Therefore, we do not use CASSCF energies as a quantitative reference.

\subsection{Additional Implementation Details}

We implemented the workflow with custom Python scripts (Python 3.11.10). We used several libraries: RDKit (2022.09.5) for cheminformatics, h5py (1.12.0)~\cite{koziol2020hdf5} for HDF5 storage, ASE (3.25.0)~\cite{hjorth2017atomic} for converting atomic numbers to element symbols, OpenBabel (3.1.0)~\cite{o2008pybel,o2011open} for format conversion, and Pysisyphus (v1.0.0)~\cite{pysisyphus} for generating redundant internal coordinates. The OM2 and OM2/MRCI calculations used the MNDO package.~\cite{MNDO2019} The CASSCF calculations used Molpro.~\cite{werner2012molpro, werner2020molpro,muramolpro} All cheminformatics analyses used RDKit.

The dataset is provided in two files: 
\textit{Final\_property.hdf5} and \textit{final\_all.csv}. \textit{Final\_property.hdf5} is a binary file that stores the geometries and energies of all compounds, including both ground-state minima and $S_0$-$S_1$ CIs.
The structure of the HDF5 file is described in Table~1. \textit{final\_all.csv} is an auxiliary file that contains InChI strings, SMILES strings, compound types, ring numbers, heavy-atom counts, and benchmark results for each molecule.

\begin{table}[!htbp]
\centering
\begin{tabular}{lllp{7cm}}
\toprule
\rowcolor{blue!20}
\textbf{No.} & \textbf{Source} & \textbf{Key} & \textbf{Description} \\
\midrule
1& Attribution & labels & Atomic Label \\
2& Ground & coordinates & Coordinates of Ground-State Minima \\
3& Ground & energy & Ground-State Minima Energies \\
4& CI & coordinates & Coordinates of Conical Intersections \\
5& CI & energy & Conical Intersections Energies \\
\bottomrule
\end{tabular}
\caption{HDF5 file structure.}
\label{tab:hdf5_structure}
\end{table}

\section{Results}

\subsection{Overview of the CI Dataset}

We selected 441,897 molecules with valid SMILES strings from the QCDGE dataset and re-optimized their geometries at the OM2 level to ensure consistency. Only 75 structures ( $ < 0.02 \%$ ) failed to converge, yielding 441,822 ground-state minima. Subsequent connectivity checks removed 6,803 structures with changes in molecular topology or connectivity. The remaining 435,019 geometries were subjected to single-point excited-state calculations at the OM2/MRCI level. Approximately 430,000 calculations converged successfully, corresponding to more than $97\%$ of the input structures. These structures were then used for CI optimization.
The CI optimizations yielded 283,645 converged geometries. The lower convergence rate at this stage can be attributed to several factors. In particular, ground-state geometries may lie far from the CI region and are therefore not always suitable starting points for CI optimization. Other possible causes include SCF convergence failures for distorted structures, mixing between active and inactive orbitals, and numerical instabilities in the coupled-perturbed equations. Such convergence difficulties are common in large-scale multiconfigurational calculations and are discussed further in the following section.
Finally, a second set of filters was applied to the optimized MECI structures, as described in Step 5. Dissociation products were removed because the dataset focuses on internal-conversion funnels rather than photodissociation pathways. We also excluded CI structures with energies below those of the corresponding ground-state minima. After these filtering steps, 259,184 molecules remained in the final QCDGE-CI dataset.

\subsection{Chemical Space Analyses}

We performed chem-informatics analyses to show the chemical diversity of the dataset. These cover elemental composition, compound types, ring distributions, functional groups, molecular shapes, bond lengths, and energy distributions.

\subsubsection{Elemental composition}

We examined the elemental composition of the molecules. The dataset has small molecules with up to ten heavy atoms (C, N, O, F). Table 2 lists the fourteen elemental compositions found. The largest group contains C, N, and O together, making up about $44.3\%$ of the total. Molecules with C and O are the second most common, followed by those with C and N. In contrast, molecules with one heteroatom (N or O), two heteroatoms (F and N, or F and O), or three heteroatoms (F, N, O) are very rare. Each of these categories has only a single-digit count. Compared with the original QCDGE dataset,~\cite{Zhu2024} the counts in Table 2 are about half, but the overall trends are very similar.

\begin{table}[!htbp]
\centering
\small
\caption{Element composition (heavy atoms) of the QCDGE-CI dataset.}
\label{tab:element_composition}
\begin{tblr}{
  width = \textwidth,
  colspec = {|X[2.0,c]|X[c]|X[c]|X[c]|X[c]|X[c]|X[c]|X[c]|X[c]|X[1.5,c]|X[1.5, c]|},
  row{1,2} = {bg=blue!20, font=\bfseries},
  row{3-Z} = {bg=white},
  hline{1,3-18} = {1-11}{solid},
  hline{2} = {2-11}{solid},
  vline{1-11} = {solid},
}
\SetCell[r=2]{c} Element Composition & \SetCell[c=10]{c} Number of Heavy Atoms & & \\
\cline{2-10}
 & 2 & 3 & 4 & 5 & 6 & 7 & 8 & 9 & 10 & Counts  \\
C & 1 & 3 & 19 & 56 & 159 & 378 & 1151 & 3319 & 2353 & 7439 \\ 
N & 1 & 2 & 1 & 2 & 2 & 0 & 0 & 0 & 1 & 9 \\ 
O & 0 & 2 & 1 & 0 & 1 & 0 & 0 & 0 & 0 & 4 \\ 
CO & 0 & 8 & 39 & 150 & 548 & 1834 & 6486 & 27113 & 19149 & 55327 \\
CN & 1 & 13 & 69 & 245 & 818 & 2119 & 5094 & 13846 & 24995 & 47200 \\ 
CF & 1 & 1 & 11 & 26 & 57 & 111 & 196 & 312 & 2112 & 28327 \\ 
NO & 1 & 4 & 3 & 9 & 7 & 2 & 2 & 1 & 0 & 29 \\ 
NF & 1 & 2 & 3 & 0 & 2 & 1 & 0 & 0 & 0 & 9 \\ 
OF & 1 & 0 & 0 & 0 & 0 & 0 & 0 & 0 & 0 & 1 \\ 
CNF & 0 & 2 & 17 & 32 & 82 & 198 & 437 & 1219 & 8725 & 10712 \\ 
CNO & 0 & 6 & 73 & 293 & 1076 & 3239 & 10567 & 42581 & 56992 & 114827 \\
COF & 0 & 2 & 11 & 44 & 104 & 215 & 403 & 767 & 7204 & 8750 \\ 
NOF & 0 & 2 & 0 & 1 & 0 & 0 & 0 & 0 & 0 & 3 \\ 
CNOF & 0 & 0 & 5 & 27 & 55 & 139 & 356 & 1282 & 10183 & 12047 \\ 
\textbf{Total} & 7 & 47 & 252 & 885 & 2911 & 8236 & 24692 & 90440 & 131714 & 259184 \\
\end{tblr}
\end{table}

\subsubsection{Compound type}

\begin{figure}[!htbp]
    \centering
    \includegraphics[width=1\linewidth]{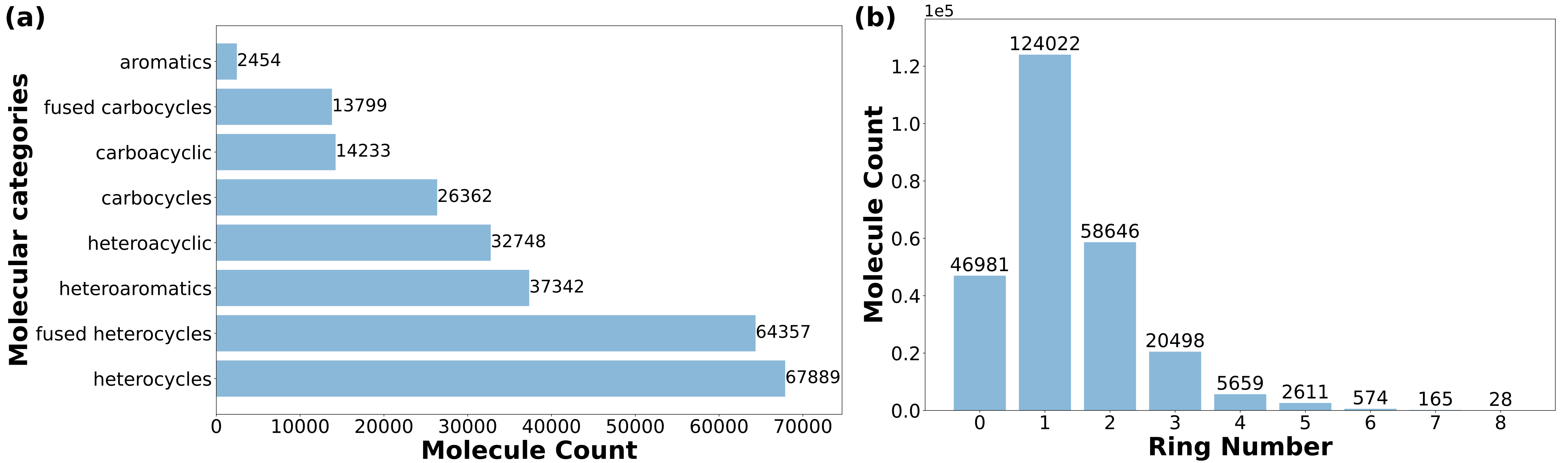}
    \caption{Analysis of compound type and molecular ring distribution in the QCDGE-CI dataset. (a) Compound type distribution. (b) Molecular ring distribution.}
    \label{fig:topology and compound type}
\end{figure}

Figure~\ref{fig:topology and compound type}(a) shows the distribution of compound types. The dataset has eight types: heterocycles (26.19\%), fused heterocycles (24.83\%), heteroaromatics (14.41\%), heteroacyclic (12.64\%), carbocycles (10.17\%), carboacyclic (5.49\%), fused carbocycles (5.32\%), and aromatics (0.95\%). Types with heteroatoms are the most common, which highlights the importance of heteroatoms in the dataset. The presence of eight types, each with a non-negligible share, shows reasonable chemical diversity. Compared with the original QCDGE dataset,~\cite{Zhu2024} the type ranking is similar. However, the drop in heteroacyclic molecules is larger than that of heteroaromatics molecules, which slightly changes their relative order.

\subsubsection{Ring number analysis}

Figure~\ref{fig:topology and compound type}(b) shows the ring number distribution.Rings are present in more than 80\% of the molecules, with approximately half containing a single ring. Although molecules with eight rings are rare, the broad range of ring counts highlights the structural diversity of the dataset.

Compared with the original QCDGE dataset,~\cite{Zhu2024} the ring distribution is similar. Most ring categories show a reduction of about 40\%, matching the overall drop in molecule count. The decrease ratio in zero-ring molecules is larger than for other categories. This is likely because acyclic molecules break more easily during CI optimization, so they were excluded from the final set.

\subsubsection{Functional group analysis}

We found 101 functional groups across the dataset. The 20 most common ones are listed in Table 3. The C=C double bond is the most abundant, while the cyclic amide (any ring size) is the least common in the top 20. Most groups contain at least one heteroatom (N, O, or F), with oxygen being the most frequent, and most also have at least one unsaturated bond. This is expected because heteroatom-containing molecules make up a large part of the original QCDGE dataset, and unsaturated molecules are more likely to form $S_0$-$S_1$ MECIs. 

We also performed a scaffold diversity analysis~\cite{bemis1996properties} to check the variety of molecular backbones. Details are given in the Supplementary information(SI).

\begin{table}[!htbp]
\centering
\resizebox{\textwidth}{!}{
\begin{tabular}{|>{\centering\arraybackslash}m{1cm}|>{\centering\arraybackslash}m{2.8cm}|>{\centering\arraybackslash}m{4.5cm}|>{\centering\arraybackslash}m{1cm}|>{\centering\arraybackslash}m{2.5cm}|>{\centering\arraybackslash}m{4.8cm}|}
\hline
\rowcolor{blue!20}
\textbf{No.} & \textbf{General structure} & \textbf{Substituents} & \textbf{No.} & \textbf{General structure} & \textbf{Substituents} \\ 
\hline
1 & \includegraphics[width=3cm,height=2.6cm,keepaspectratio]{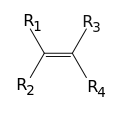} & any compound with a double bond & 11 & \includegraphics[width=2.2cm,height=1.8cm,keepaspectratio]{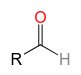} & any compound with aldehyde \\ 
\hline
2 & \includegraphics[width=3cm,height=2.6cm,keepaspectratio]{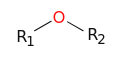} & R1, R2=alkyl  & 12 & \includegraphics[width=1.8cm,height=1.4cm,keepaspectratio]{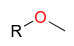} & any compound with methyl ether \\ \hline
3 & \includegraphics[width=2.2cm,height=1.8cm,keepaspectratio]{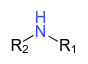} & R1, R2=alkyl & 13 & \includegraphics[width=2cm,height=1.6cm,keepaspectratio]{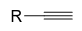} & any compound with a triple bond  \\ 
\hline
4 & \includegraphics[width=1.5cm,height=0.9cm,keepaspectratio]{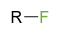} & R=acyl  & 14 & \includegraphics[width=2.6cm,height=2.2cm,keepaspectratio]{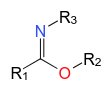} & \makecell{R1=H, alkyl \\ R2=alkyl \\ R3=H, alkyl,aryl}  \\ 
\hline
5 & \includegraphics[width=2cm,height=1.6cm,keepaspectratio]{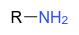} & R=alkyl  & 15 & \includegraphics[width=2.8cm,height=2.4cm,keepaspectratio]{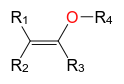} & \makecell{R1, R2, R3=H, acyl, alkyl \\  R4=H, alkyl}   \\ 
\hline
6 & \includegraphics[width=2.4cm,height=2.0cm,keepaspectratio]{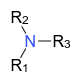} & R1, R2, R3=alkyl  & 16 & \includegraphics[width=3cm,height=2.6cm,keepaspectratio]{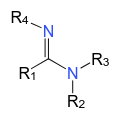} & R1, R2, R3, R4=H, alkyl  \\ 
\hline
7 & \includegraphics[width=2.4cm,height=2.0cm,keepaspectratio]{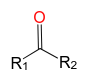} & R1, R2=alkyl, aryl  & 17 & \includegraphics[width=2.4cm,height=2cm,keepaspectratio]{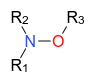} & R1, R2, R3=H, alkyl  \\ 
\hline
8 & \includegraphics[width=2.4cm,height=2cm,keepaspectratio]{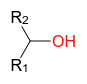} & R1, R2=alkyl  & 18 & \includegraphics[width=2.4cm,height=2cm,keepaspectratio]{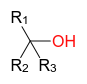} & R1, R2, R3=H, alkyl  \\ 
\hline
9 & \includegraphics[width=2.4cm,height=2cm,keepaspectratio]{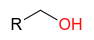} & R=alkyl, aryl  & 19 & \includegraphics[width=2.2cm,height=1.8cm,keepaspectratio]{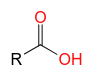} & any compound with carboxyl  \\ \hline
10 & \includegraphics[width=2.2cm,height=1.8cm,keepaspectratio]{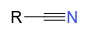} & any compound with a cyanide  & 20 & \includegraphics[width=2.8cm,height=2.4cm,keepaspectratio]{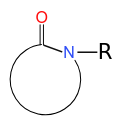} & R=H, alkyl, aryl  \\ \hline
\end{tabular}
}  
\caption{General structural motifs and substituent moieties of the top 20 functional groups in the QCDGE-CI dataset. The InDraw package was used to draw general structural motifs.}
\label{tab:functional group}
\end{table}

\subsubsection{PMI analysis}

\begin{figure}[!htbp]
    \centering
    \includegraphics[width=1\linewidth]{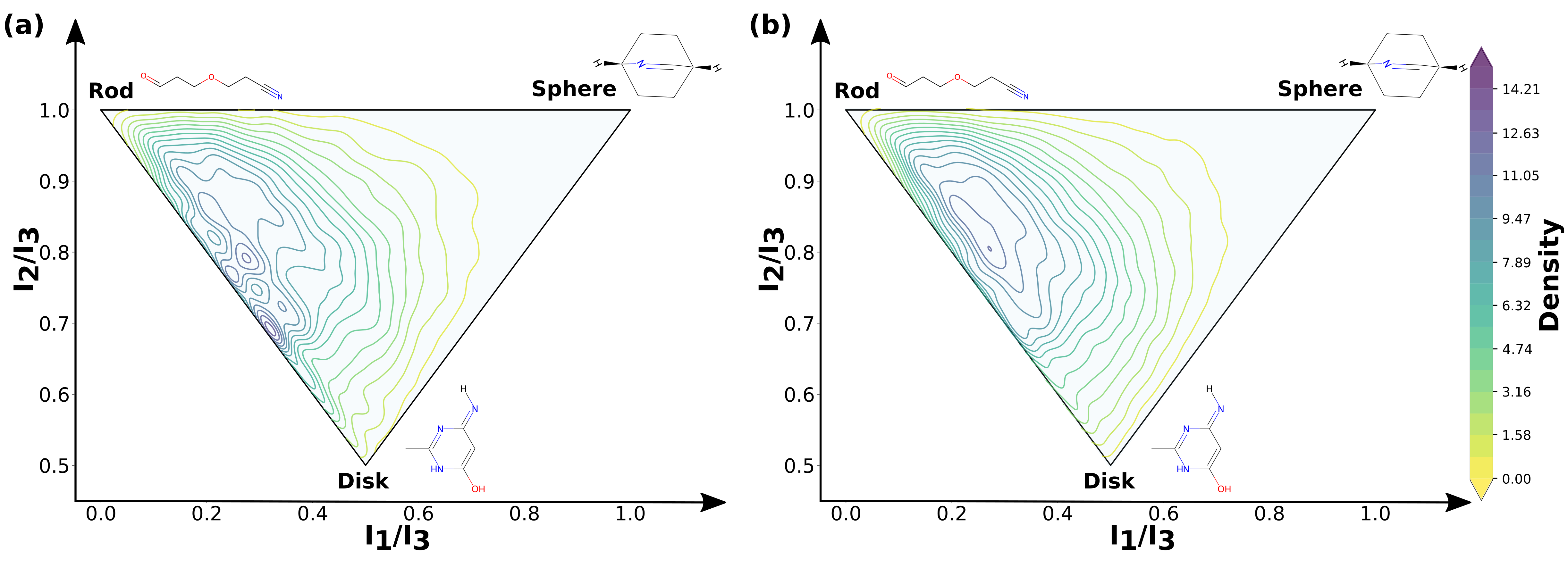}
    \caption{Analysis of principal moments of inertia in the QCDGE-CI dataset. Here $\mathbf{I_1, I_2, I_3}$ are defined as the principle moments of inertia with order of $\mathbf{I_1} <  \mathbf{I_2} < \mathbf{I_3}$.  The x-axis and y-axis are normalized moments of inertia. (a) Ground-state geometries. (b) CI geometries.}
    \label{fig:pmi_analysis}
\end{figure}

To study molecular shape diversity, we performed a principal moments of inertia (PMI) analysis~\cite{Sauer2003} for both ground-state and 
$S_0$-$S_1$ CI geometries. The resulting shape distributions are shown in Figure~\ref{fig:pmi_analysis}.

The dataset covers many structural types for both ground and MECI geometries. Most molecules lie between the rod-like and disc-like regions, suggesting extended or moderately planar shapes. Few molecules are spherical. The shape of disc-like molecules is clearly lower at the CI than in the ground-state set. This decrease is likely attributable to the substantial structural distortions required to access the CI region, encompassing bond elongation, torsional deformation, and ring puckering. Collectively, these distortions give rise to more three-dimensional molecular architectures.

\subsubsection{Bond length analysis}

\begin{figure}[!htbp]
     \centering
     \includegraphics[width=1\linewidth]{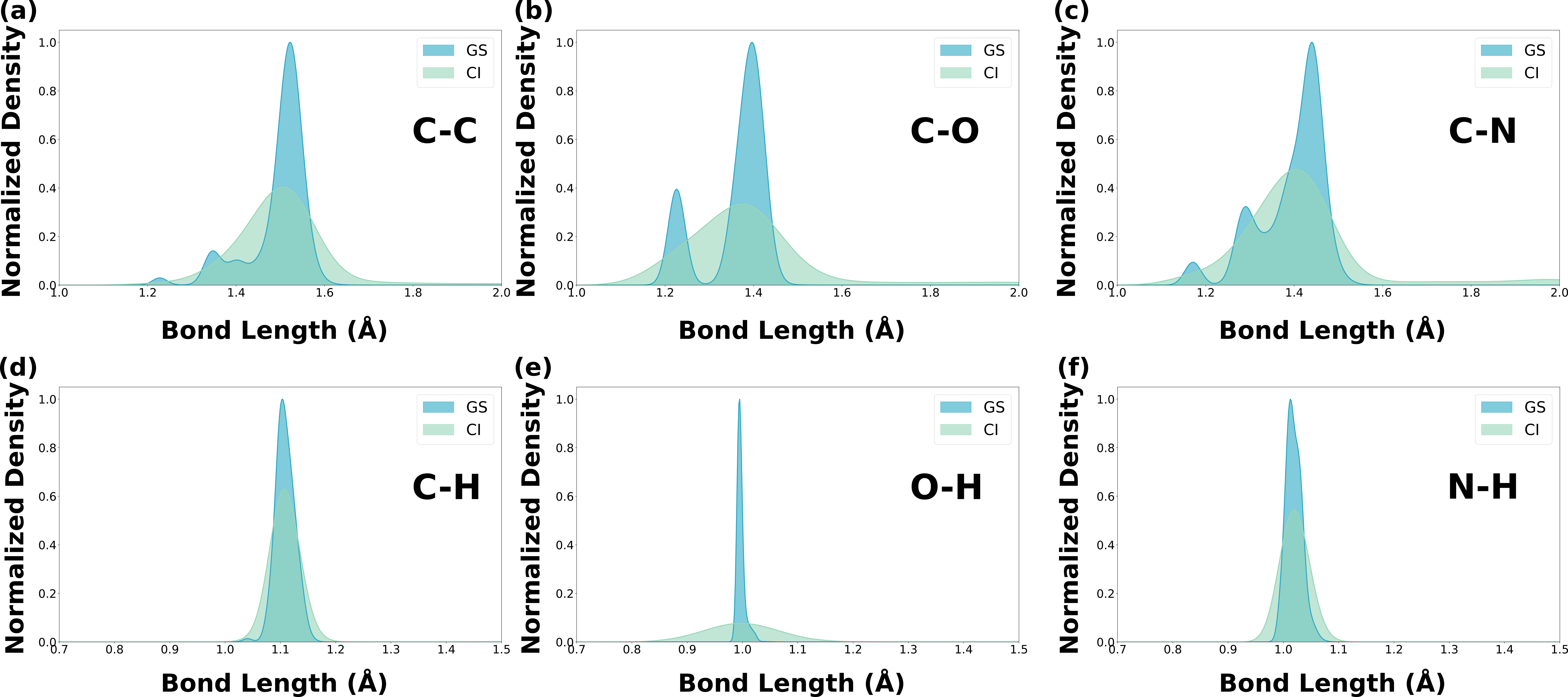}
     \caption{Bond length distributions of (a) C-C bonds, (b) C-O bonds, (c) C-N bonds, (d) C-H bonds, (e) O-H bonds, (f) N-H bonds. Here GS and CI refer to ground-state minima and CIs}
     \label{fig:bond_length_analysis}
 \end{figure}

We analyzed the bond-length distributions for C-C, C-O, C-N, C-H, O-H, and N-H bonds. Because some bonds may break during relaxation from the ground state to the CI, we identified the selected bonds in each ground-state minimum and tracked their corresponding distances in the MECI structure. Several of these bonds became substantially elongated at the CI, suggesting bond weakening or dissociation. The resulting distributions are shown in Figure~\ref{fig:bond_length_analysis}; additional details are provided in the SI.

The distributions of the three bond types (C-C, C-O, and C-N) at the ground-state minima are relatively broad, with most values ranging from 1.25 \AA\ to 1.75 \AA . Distinct fine-resolved peak structures appear, corresponding to different bond orders. At the CI, these distributions become broader, and some bond distances appear between these peaks. As shown in Figure~S3, some C=C double bonds are elongated at the $S_0$-$S_1$ MECI structures. 
Among all types of bonds, the aromatic C-O bond length at the MECI shows the most gradual distribution. 
The distributions of single bonds involving H atoms also become broader from the ground-state minima to the CIs.   

\subsubsection{Energy analysis}

The chemical diversity of the QCDGE-CI dataset also appears in the energy distributions. These include the 
the $S_1$ vertical excitation energies (VEEs) of the ground-state minima, and the 
energies at $S_0$-$S_1$ MECIs with respect to their corresponding ground-state minima. 

\begin{figure}[!htbp]
    \centering
    \includegraphics[width=1\linewidth]{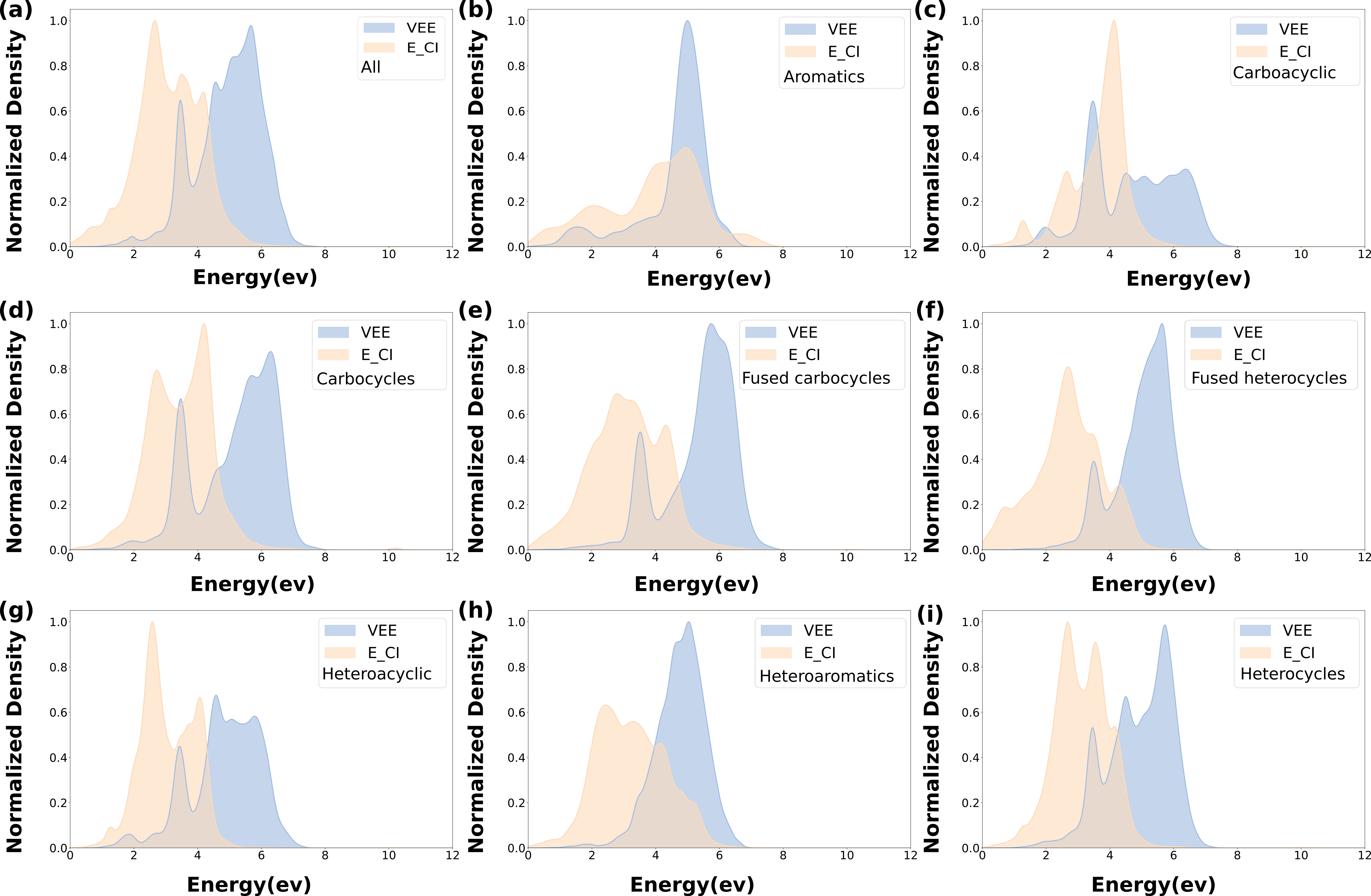}
    \caption{Distribution of $S_1$ VEE and $S_0$-$S_1$ CI energies in the QCDGE-CI dataset. (a) All molecules. (b) Aromatics. (c) Carboacyclic. (d) Carbocycles. (e) Fused carbocycles. (f) Fused heterocycles. (g) Heteroacyclic (h) Heteroaromatics. (i) Heterocycles. }
    \label{fig:energy_analysis}
\end{figure}

As shown in Figure~\ref{fig:energy_analysis}, the $S_1$ VEEs of the ground-state minimum are distributed from 1 eV to 8 eV.
For most categories, the energy of the $S_0$-$S_1$ MECIs are lower than the $S_1$ VEEs at the ground-state minima.

Among the molecular classes, aromatics have the widest spread of $S_0$-$S_1$ CI energies. 
Carboacyclic compound shows relatively narrower $S_0$-$S_1$ CI energy ranges. 
Fused heterocycles and fused carbocycles display intermediate CI energy distributions. 

Regarding the $S_1$ VEEs, 
carboacyclic, carbocycles, and fused carbocycles compounds have broader ranges. 
Aromatics and heteroaromatics are mostly confined to lower
$S_1$ VEE values.
The relative energy distribution is also shown in Figure~S4.

\subsection{Benchmark Results}

In this section, we present our benchmark results, focusing on two types of comparisons. The first compares our ground-state minima against DFT reference values, while the second concerns the CI structures and has two parts. First, we compare the CI geometries from the (4,4) and (8,8) active spaces at the OM2/MRCI level. This helps us see how sensitive the structures are to the active-space choice. Second, we benchmark the OM2/MRCI(4,4) geometries against CASSCF results on a representative subset. This provides a high-level reference for the semi-empirical method.

\subsubsection{Ground-state minima: comparing with DFT results}

\begin{figure}[!htbp]
    \centering
    \includegraphics[width=1\linewidth]{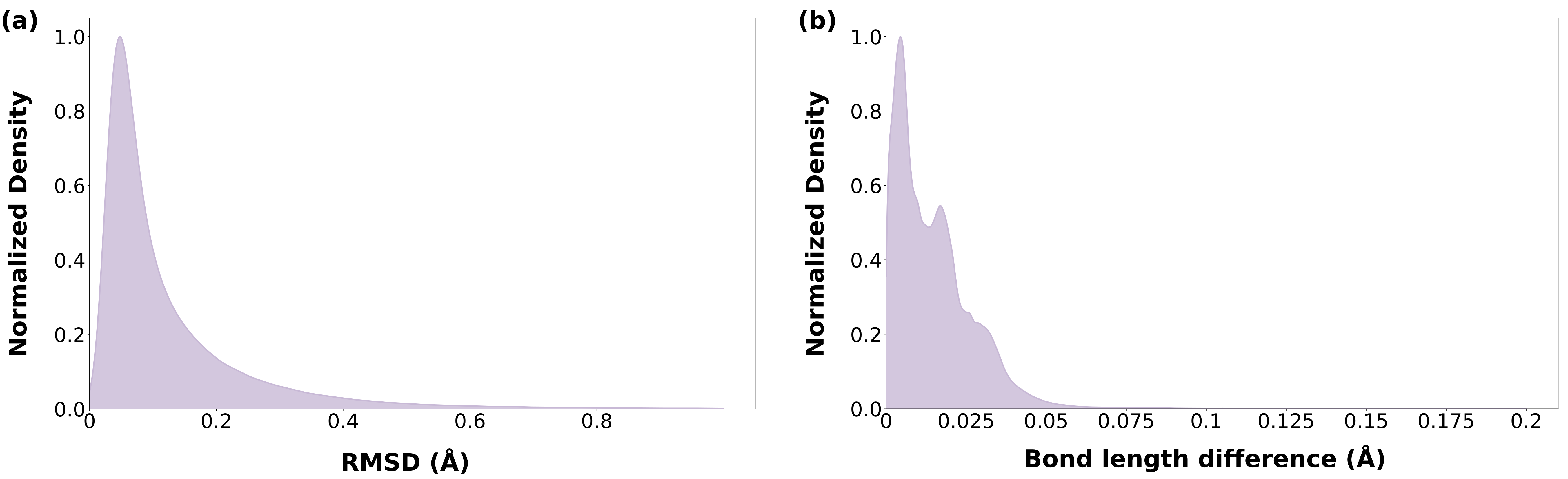}
    \caption{Distribution of benchmark results. (a) RMSD comparison of different active space. (b) Bond length comparison of Ground state minia.}
    \label{fig:ground statebenchmark results}
\end{figure}

We validated the final 259,184 OM2-optimized geometries by comparing them with their DFT-optimized counterparts from the QCDGE dataset.
As discussed above, we focused on bond-length comparisons and RMSD values. The comparison (Figure~\ref{fig:ground statebenchmark results}) shows excellent agreement between the two sets of geometries, with a mean absolute error (MAE) of only $0.015$ \AA\ for bond-lengths and $0.127$ \AA\ for RMSD values.
This confirms that the ground-state geometries in the final QCDGE-CI dataset are reliable.

\subsubsection{CI geometries: comparison with the enlarged active space}

\begin{figure}[!htbp]
    \centering
    \includegraphics[width=1\linewidth]{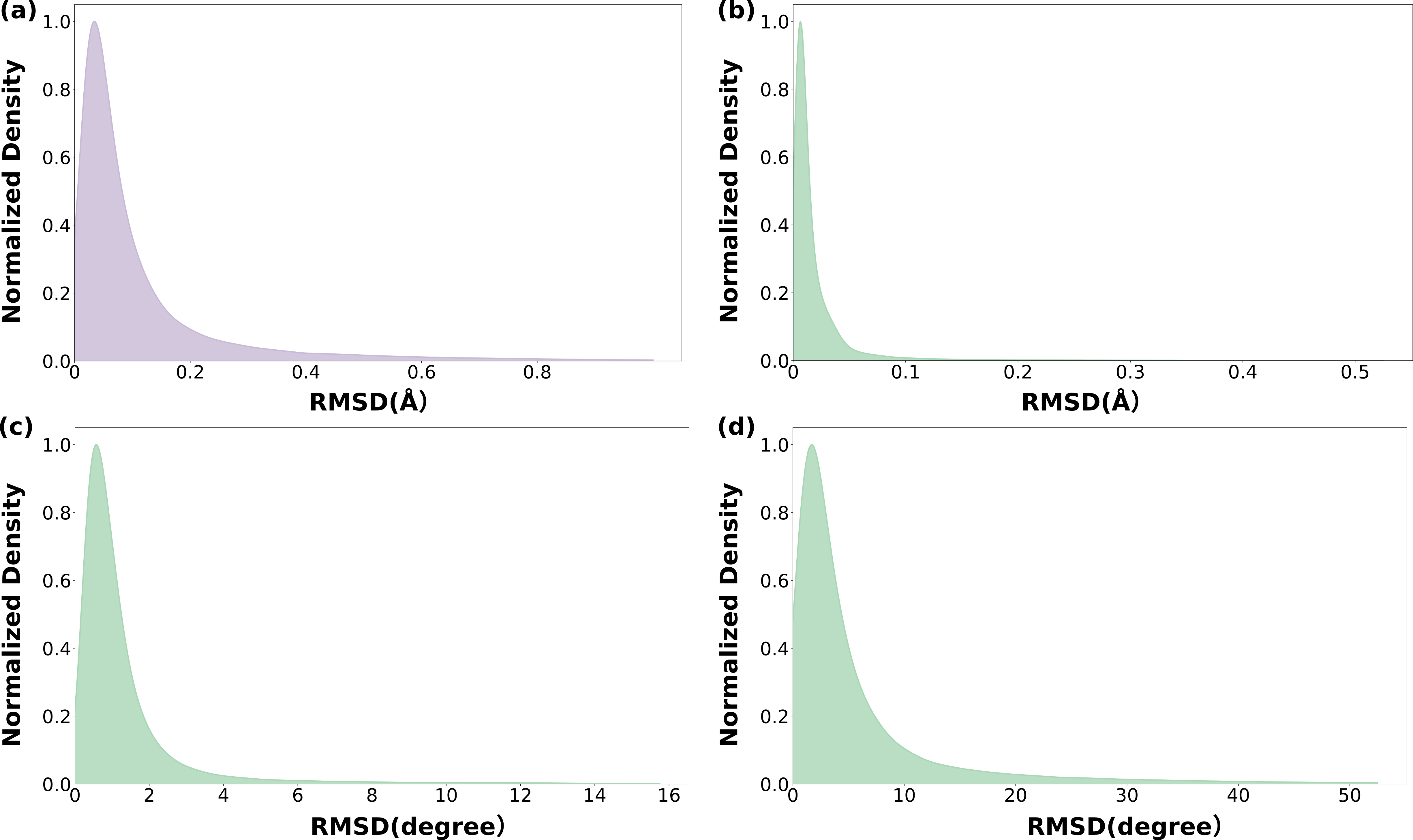}
    \caption{Distribution of CI state geometry benchmark results in larger active spaces. (a) Overall RMSD benchmark results of CI geometries. (b) Bond RMSD benchmark results of CI geometries. (c) Bend RMSD benchmark results of CI geometries. (d) Dihedral RMSD benchmark results of CI geometries.}
    \label{fig:large active space benchmark results}
\end{figure}

Starting from the (4,4) geometries, we re-optimized all structures with an enlarged (8,8) active space at the same OM2/MRCI level. Of these calculations, 191,094 converged successfully, and the two sets of MECI geometries were then compared. As shown in Figure~\ref{fig:large active space benchmark results}(a), the overall agreement is reasonably good, with a mean absolute error (MAE) of 
$0.121$ \AA. 
This suggests that (4,4) active space yields structures comparable to those obtained with the larger active space.

To assess whether this agreement varies across different molecular classes, we computed the average RMSD for each of the eight compound types (Table~\ref{tab:RMSD_8}). 
The results show that the deviations between the (4,4) and (8,8) geometries are similar across all types. This consistence indicates that the active-space dependence does not exhibit a strong bias toward any particular molecular class.

To further understand the origin of the observed deviations, we decomposed the RMSD into contributions from bond lengths, bond angles, and dihedral angles using redundant internal coordinates, as shown in Figure~\ref{fig:large active space benchmark results}(b)-(d) and Table~\ref{tab:RMSD_8}.

The average difference in dihedral angles between the (4,4) and (8,8) geometries is about 8 degrees. This deviation can arise from two distinct sources. One is single-bond torsion, which leads to different conformations around the same CI structure. The other is the intrinsic extension of the CI seam along torsional coordinates, where the (4,4) and (8,8) calculations may converge to different points on the same seam. In both cases, the reported MECI should be viewed as a representative point on the seam. This limitation does not affect the overall reliability of the dataset for large-scale screening and machine-learning applications. For studies that require precise torsional geometries, the MECI structures reported here provide a useful starting point for further refinement.

\begin{table}[htbp]
\centering
\caption{Benchmark comparison between the current dataset and OM2/MRCI(8,8). Counts denote the number of compounds that converged at the OM2/MRCI(8,8) level. The table reports the averaged RMSD values for overall geometry, bond lengths, bond angles, and dihedral angles. }
\label{tab:RMSD_8}
\begin{tabular}{lrrrrr}
\rowcolor{blue!20}
\toprule
CompoundType & Count & RMSD(\AA) & Bond(\AA) & Bend(degree) & Dihedral(degree) \\
\midrule
carboacyclic & 9346 & 0.155 & 0.012 & 1.343 & 8.021 \\
heteroacyclic & 23923 & 0.172 & 0.017 & 1.167 & 8.619 \\
carbocycles & 18766 & 0.119 & 0.015 & 1.266 & 5.823 \\
heterocycles & 51024 & 0.121 & 0.021 & 1.411 & 6.374 \\
heteroaromatics & 27114 & 0.127 & 0.028 & 1.720 & 6.290 \\
fused carbocycles & 9906 & 0.082 & 0.018 & 1.363 & 4.180 \\
fused heterocycles & 49213 & 0.093 & 0.024 & 1.534 & 4.897 \\
aromatics & 1802 & 0.163 & 0.031 & 2.217 & 8.567 \\
\bottomrule
\end{tabular}
\end{table}

\subsubsection{CI geometries: comparison with CASSCF results}

\begin{figure}[!htbp]
    \centering
    \includegraphics[width=1\linewidth]{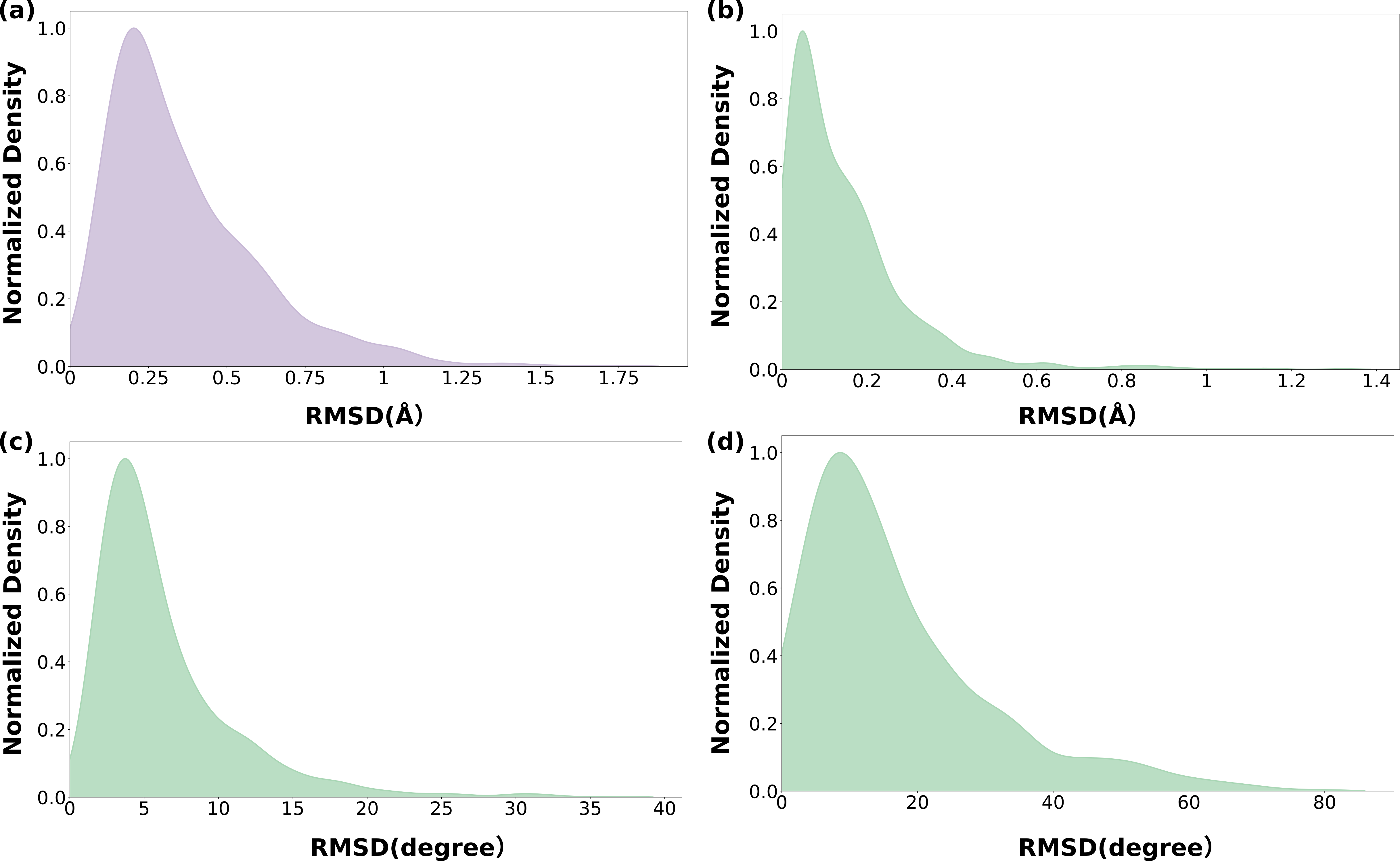}
    \caption{Distribution of CI state geometry benchmark results in CASSCF method. (a) Overall RMSD benchmark results of CI geometries. (b) Bond RMSD benchmark results of CI geometries. (c) Bend RMSD benchmark results of CI geometries. (d) Dihedral RMSD benchmark results of CI geometries.}
    \label{fig:CASSCF method benchmark results}
\end{figure}

For each of the eight compound types, we randomly selected 200 CI geometries and re-optimized them at the CASSCF level using two different active spaces, (4,4) and (8,8). The convergence ratios were similar across all types, ranging between 0.76 and 0.88, and the average RMSD values fell within a narrow range of $0.297$-$0.399$ \AA\ (Table~\ref{tab:RMSD_CAS}).
Notably, neither the convergence ratios nor the RMSD values depend strongly on the active-space choice. Consistent with the observations in the previous subsection, the main differences between the OM2/MRCI and CASSCF geometries origniate from torsional motions, rather than backbone deformations, as illustrated in Figure~\ref{fig:CASSCF method benchmark results} and Table~\ref{tab:RMSD_CAS}.

Overall, these benchmarks reinforce our confidence in the OM2/MRCI(4,4) geometries and support using this dataset for large-scale photochemical studies.

\begin{table}[htbp]
\centering
\caption{Benchmark comparison between the current dataset and CASSCF level. Counts denote the number of compounds that converged at the CASSCF level. The table reports the averaged RMSD values for overall geometry, bond lengths, bond angles, and dihedral angles.}
\label{tab:RMSD_CAS}
\begin{tabular}{lrrrrr}
\rowcolor{blue!20}
\toprule
CompoundType & Count & RMSD(\AA) & Bond(\AA) & Bend(degree) & Dihedral(degree) \\
\midrule
carboacyclic & 160 & 0.357 & 0.086 & 5.577 & 17.042 \\
heteroacyclic & 165 & 0.391 & 0.127 & 4.621 & 17.628 \\
carbocycles & 152 & 0.372 & 0.170 & 6.201 & 15.455 \\
heterocycles & 155 & 0.397 & 0.149 & 6.210 & 18.658 \\
heteroaromatics & 176 & 0.399 & 0.200 & 8.016 & 16.977 \\
fused carbocycles & 159 & 0.297 & 0.156 & 6.489 & 14.535 \\
fused heterocycles & 161 & 0.324 & 0.163 & 6.482 & 16.270 \\
aromatics & 176 & 0.367 & 0.124 & 6.579 & 19.076 \\
\bottomrule
\end{tabular}
\end{table}

\section{Discussions}

We have constructed a large-scale dataset of ground-state and $S_0$-$S_1$  MECI geometries, analyzed its chemical space coverage, and validated its reliability through benchmark calculations. The key aspects of this work are discussed below.

Our goal is to construct a large-scale dataset for statistical and machine-learning applications, not to provide benchmark-quality data for individual molecules. The OM2/MRCI method was chosen for its balance of efficiency and accuracy: it is fast enough to process hundreds of thousands of molecules, provides the gradients and coupling vectors needed for CI optimization, and has been validated in previous studies. The fixed (4,4) active space, though an approximation, ensures consistency across the dataset and avoids the impracticality of varying active spaces per molecule. Benchmark comparisons with enlarged active space calculations support the validity of this pragmatic choice.

The comparison with enlarged (8,8) active space calculations was designed to assess whether the (4,4) geometries are stable with respect to active-space expansion. The results show reasonable agreement and the deviations are not concentrated in any particular molecular class. Further decomposition of the RMSD reveals that the differences arise primarily from torsional motions of flexible fragments, while the molecular backbones remain largely unchanged. This suggests that the (4,4) active space captures the essential structural features of the CI geometries.

The CASSCF benchmark provides a high-level reference for the semi-empirical results. The OM2/MRCI geometries show reasonable agreement with CASSCF-optimized structures across all compound types. This comparison confirms that the semi-empirical method yields qualitatively reliable geometries suitable for large-scale applications.

The reasonable agreement with CASSCF demonstrates the practical value of the OM2/MRCI approach. However, this success applies only to the subset of molecules for which the CI optimization converged. For the remaining molecules, the optimization failed and this failure is itself a critical piece of information. We therefore discuss the causes of these convergence difficulties in detail below.

CI optimization itself is more challenging than ground-state optimization, and understanding its difficulties is important for interpreting the dataset. The primary goal of this work is to construct a diverse and statistically meaningful dataset, not to systematically solve convergence problems. Nevertheless, this issue deserves some discussion because it affects a non-negligible fraction of the calculations and provides useful context for users of the dataset.

In our calculations, over 80\% of the unconverged OM2/MRCI(4,4) jobs reached a region where the energy gap between the two electronic states became very small and the geometry changes were no longer significant, yet the calculation still did not fully converge. A key reason is the discontinuity of the electronic wavefunction. In methods that rely on active-space selection, such as OM2/MRCI and CASSCF, the wavefunction is represented by a set of orbitals partitioned into active and inactive subspaces.~\cite{Shepard_acp_1987} 
As molecular geometry changes, the coefficients of the orbitals undergo re-mixing.~\cite{Shepard_acp_1987} This re-mixing is not a simple exchange between active-space and external orbitals; it involves a redistribution of orbital characters that can vary from one geometry to the next. In active-space methods, only the orbitals inside the active space are used to generate electronic excitations. Therefore, even when the geometry changes only slightly, this re-mixing can prevent the wavefunction from becoming a smooth function of nuclear coordinates. This directly affects the energy gradients and nonadiabatic coupling vectors. Such ill-defined or numerically unstable issues easily lead to convergence failures in the CI optimization.

Another complication arises from the nature of the CI optimization itself. CI optimization aims to locate a minimum on the crossing seam, which differs fundamentally from ground-state optimization. It requires satisfying two conditions simultaneously: energy degeneracy between the two electronic states and a constrained minimum on the upper state. This makes the optimization landscape more complex. In addition, the potential energy profiles of excited states are also very complicated.

For the same reasons discussed above, enlarging the active space to (8,8) can resolve some cases, but it may also introduce new orbital mixing and lead to other convergence problems. In fact, most of these (8,8) calculations followed a similar pattern: they reached the CI seam region but still failed to converge fully. 

Although CASSCF optimizes both orbitals and CI coefficients simultaneously and allows fractional occupations, which makes SCF convergence easier, we still observed energy oscillations and sudden jumps after several steps. Many of these CASSCF calculations also reached the CI seam region but did not fully converge. Changing the active space sometimes improved convergence, but it could also cause previously converged cases to fail. These convergence issues are not specific to our dataset; they are common in multi-configurational calculations across diverse molecular systems.

For a dataset of this scale, performing multiple CI optimizations from different starting points, or exploring the seam space with more advanced sampling techniques, is computationally prohibitive even at the semiempirical level. We therefore adopted a pragmatic strategy: each molecule includes a single MECI structure obtained from one starting geometry. This is clearly a limitation. A molecule may have several distinct minima on the CI seam, and different starting structures may lead to different CI geometries. This issue is not addressed in the current dataset and remains a direction for future work.

This dataset is intended for qualitative to semi-quantitative applications. 
It is suitable for machine learning training, statistical analysis, and preliminary screening of CI properties across diverse chemical spaces. 
It is not designed to replace high-level multi-reference calculations for individual molecules where high accuracy is required. 
Users should check the convergence flags and RMSD values in the dataset to assess the reliability of specific entries before using them in downstream work. 
With over 259k CI structures, the dataset provides a statistically significant resource. We hope it will serve as a useful starting point for exploring photochemical reaction patterns and for developing data-driven models of excited-state chemistry.

\section {Conclusions} 

In this work, we constructed the QCDGE-CI dataset, a large-scale resource containing ground-state minima and $S_0$-$S_1$ conical-intersection geometries for over 259k small molecules. The dataset was generated with the OM2/MRCI method using a fixed (4,4) active space, which provides a practical balance between efficiency and accuracy for large-scale screening.
Systematic benchmarks were performed using an enlarged (8,8) active space at the same OM2/MRCI level and through CASSCF re-optimizations of representative subsets. The comparison between the (4,4) and (8,8) results assesses the robustness of the geometries with respect to the active-space size, while the CASSCF benchmarks show that the (4,4) geometries agree well with those obtained using the higher-level method. This supports their use in statistical and machine-learning applications. Comprehensive chem-informatics analyses, covering elemental composition, compound types, ring distributions, functional groups, molecular shapes, bond lengths, energy distributions and molecular backbone, confirm that the dataset covers a broad and diverse chemical space.

This dataset addresses a critical gap in the field, as comprehensive CI databases for data-driven approaches have been largely unavailable. 
By providing a diverse set of MECI structures across a wide chemical space, the QCDGE-CI dataset serves as a valuable resource for exploring structure-property relationships in excited-state chemistry. It enables large-scale pattern analysis, machine learning model training, and preliminary screening of photochemical properties.
We expect this dataset to enable new insights into photochemical mechanisms and to accelerate predictive modeling of nonadiabatic processes.

\section{Data Availability}

All data files of the QCDGE-CI dataset, are available in the Figshare repository and on the project website (\url{http://langroup.site/QCDGE_CI}). 
Here, we also provide a Python script named \textit{extract\_data\_from\_QCDGE\_CI.py} to extract data from the HDF5 file. 
The function \textit{extract\_data\_from\_QCDGE\_CI()} can be imported from this script.

The expanded-active-space OM2/MRCI calculations were performed solely for the purpose of benchmarking the (4,4) protocol. The resulting RMSD values and convergence statistics are reported in Table~\ref{tab:RMSD_8} and Figure~\ref{fig:large active space benchmark results} to quantify the structural uncertainty of the main dataset. The expanded-active-space structures are not included in the main dataset because, although calculations were performed for all 259k molecules, not all of them converged successfully, and the convergence rates varied across molecular classes. To avoid introducing data incompleteness and potential selection bias, we do not deposit these benchmark structures. Users interested in higher-accuracy structures for specific molecules are encouraged to perform their own expanded-active-space calculations using the provided (4,4) structures as starting points

\section{Code Availability}

The code for extracting data from the HDF5 file and the code used for technical validation are available in the GitHub repository: \url{https://github.com/coffee-zzh/QCDGE_CI.git}.
Additionally, the code for automated calculation submission is available in the previous work~\cite{Zhu2024}.

\section{Author Contributions}

Jiahui Zhang: Conceptualization, Data curation, Software, Formal analysis, Investigation, Methodology, Visualization, Writing-review \& editing. \\
Yifei Zhu: Conceptualization, Software, Methodology, Writing-review \& editing. \\
Chuqiao Feng: Data curation, Software, Writing-review \& editing. \\
Yingjin Ma: Software, Methodology, Writing-review \& editing. \\
Chao Xu: Conceptualization, Supervision, Project administration, Resources, Funding acquisition, Writing-review \& editing. \\
Zhenggang Lan: Conceptualization, Supervision, Project administration, Resources, Funding acquisition, Writing-review \& editing. \\

\section{Conflicts of interest}

There are no conflicts to declare.

\section{Acknowledgements}

The authors express sincere thanks to the National Natural Science Foundation of China (No. 22333003, 22361132528, 22573036 and 22673132) for financial support.
Y. Ma also thanks the support of the Project of Ganjiang Innovation Academy (Grant No. E455F001), and the helps from the supporting team of “ORISE” HPC clusters.

\section{Additional information}

Supplementary information: The online version contains supplementary material available at xxxx.
Additional details are provided, including: checking process, scaffold analysis, bond length analysis of different bond orders, relative energy analysis and additional comparisons with CASSCF result.

\bibliography{QCDGE_CI}

\end{document}